\documentclass{ptephy_v1}

\usepackage{physics}
\usepackage{amsmath}
\usepackage{bm}
\usepackage[whole]{bxcjkjatype} 

\begin{document}

\title{Floquet engineering of electric polarization with two-frequency drive}


\author{Yuya Ikeda}
\author[1]{Sota Kitamura}
\author[1,2]{Takahiro Morimoto}
\affil{Department of Applied Physics, The University of Tokyo, Hongo, Tokyo, 113-8656, Japan}
\affil[2]{JST, PRESTO, Kawaguchi, Saitama, 332-0012, Japan}


\begin{abstract}%
Electric polarization is a geometric phenomenon in solids and has a close relationship to the symmetry of the system. Here we propose a mechanism to dynamically induce and manipulate electric polarization by using an external light field. Specifically, we show that application of bicircular lights (BCLs) control the rotational symmetry of the system and can generate electric polarization. To this end, we use Floquet theory to study a system subjected to a two-frequency drive. We derive an effective Hamiltonian with high frequency expansions, for which the electric polarization is computed with the Berry phase formula. We demonstrate the dynamical control of polarization for a one-dimensional SSH chain, a square lattice model, and a honeycomb lattice model.
\end{abstract}

\subjectindex{A57 Nonequilibrium steady states;
I84 Ultrafast phenomena;
I92 Graphene, fullerene}

\maketitle

\section{Introduction}

Symmetry and topology play a central role in recent studies of condensed matter physics \cite{hasan-kane-rmp,qi-zhang-rmp,ryu-rmp}. Topological phases are characterized by nontrivial phases of Bloch wave functions in solids and host gapless excitations at the surface. Quantum Hall effect is a canonical example of a topological phase, where the Chern number defined from the Berry curvature of Bloch wave functions counts the number of chiral edge states and gives the quantized value of Hall conductance \cite{TKNN}. 
Charge pumping in 1D inversion broken systems is another topological phenomenon characterized by the Chern number \cite{thouless83}. In charge pumping, electrons are pumped through the bulk by changing the parameter of the system in a nontrivial way. Its topological characterization is given by the Chern number and the Berry curvature defined within the 2D space with the momentum and pumping parameter. In this case, the Berry curvature has a meaning of the polarization current flowing through the bulk. 
A closely related phenomenon is electric polarization in polar crystals. Electric polarization is a geometrical quantity described by the Berry phase of the Bloch wave functions, which is known as the modern theory of electric polarization \cite{resta-RMP,vanderbilt-kingsmith}. Specifically, a position operator becomes ill-defined in crystals with a periodic structure, which necessitates an alternative formulation of polarization within the momentum space. In the momentum space, the Berry connection plays a role of the position expectation value for a wave packet. Thus a Berry phase, i.e., an integral of the Berry connection, gives a good description of electric polarization. 

Polar crystals host various interesting phenomena that arise from the nontrivial geometry of Bloch wave functions. In addition to the electric polarization stated above, these include a photovoltaic effect called shift current \cite{Baltz-Kraut,Sipe,Morimoto-Nagaosa16,Nagaosa-review20}, and nonreciprocal transport in quantum tunneling \cite{kitamura-cp20,kitamura-prb20,takayoshi20}. Both effects are characterized by Berry connection (so called shift vector specifically) and have a close relationship to the modern theory of electric polarization. 
In polar crystals, control of polarity inevitably requires changing crystal structure with different compositions. Another interesting platform for polar materials is van der Waals heterostructures.  They support interfacial structures with a wide variety of combination of 2D materials and leads to high controllability of symmetry of the system including polarity \cite{Akamatsu}. However, both approaches involve changing crystal structure or interfacial structures to control polarity, which usually requires fabricating new samples and hence is costly. 
It is highly desirable if one can control the polarity of the electronic system without changing crystal structures. One promising route is utilizing an external light field to manipulate electronic structures, which is known as Floquet engineering \cite{Oka-Kitamura19,Rudner2020,andre-RMP}.

Floquet engineering relies on Floquet theory which describes nonequilibrium steady states of periodically driven systems. Floquet theory is an analog of Bloch's theorem for spatially periodic systems~\cite{Shirley1965,Sambe1973}. It enables analyses of driven systems with an effective band theory with a Floquet Hamiltonian~\cite{Casas2001,Mananga2011,Bukov2015,Eckardt2015,Mikami16}. 
Floquet engineering is a concept to create the desired state by engineering a Floquet Hamiltonian with suitable driving. It offers dynamic control of quantum phases of matter without changing the underlying chemical compositions. For example, Floquet topological phases~\cite{Oka2009,Lindner2011,Wang2014,Zhang2016} are actively studied since their topology can be controlled by driving and sometimes have no counterpart in the equilibrium~\cite{Rudner2013,Roy2017,morimoto-time-glide17,Higashikawa2019,Hu2020}.

\begin{figure}[t]
    \begin{center}
        \includegraphics[width=\linewidth]{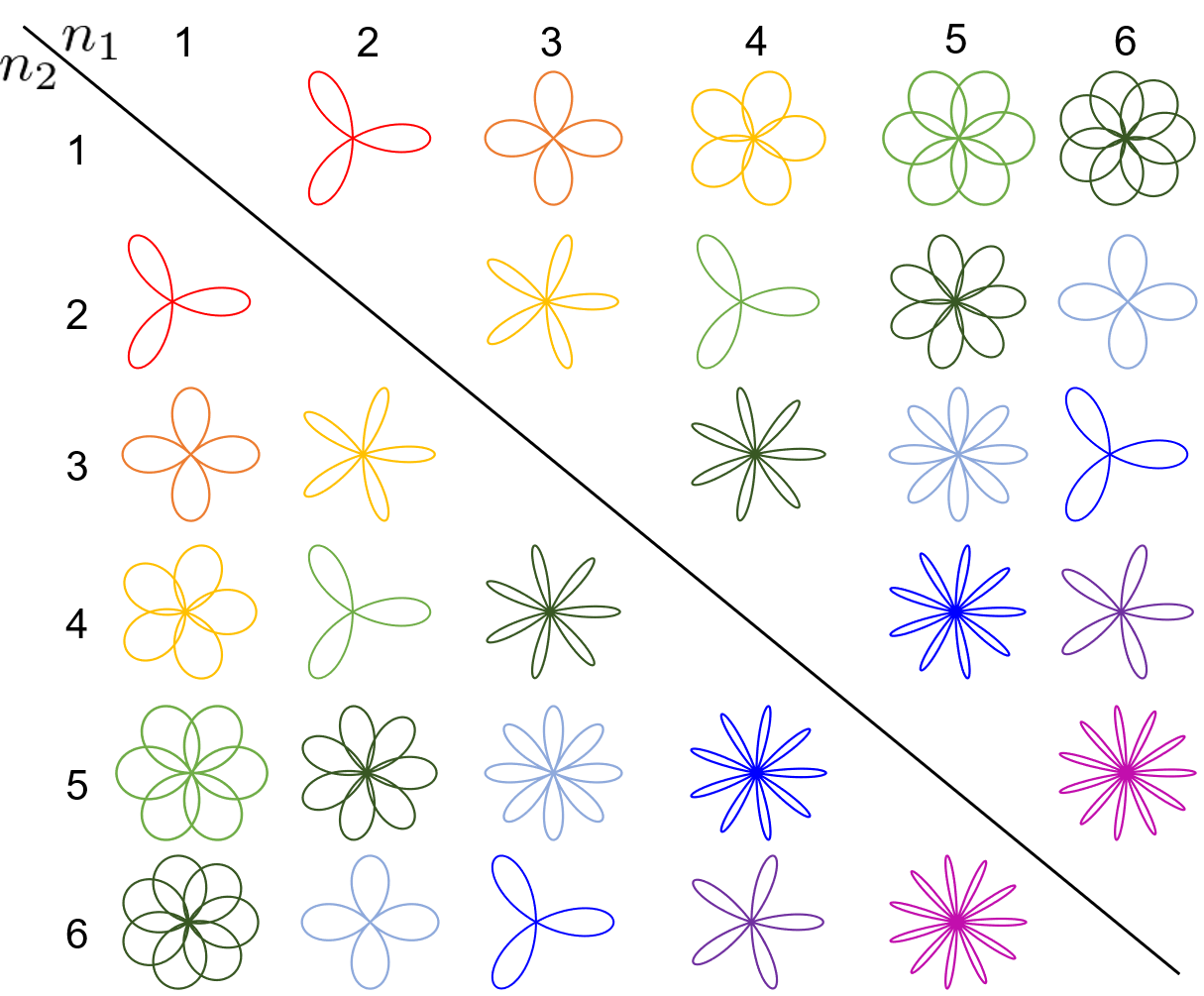}
        \caption{Rose curves as trajectories of electric fields of the bicircular light (BCL). We show the curves defined with the pairs of the integers ($n_1,n_2$) up to 6.  The relative phase is chosen as $\theta=0$. The rose curves possess $(n_1+n_2)/\mathrm{gcd}(n_1,n_2)$-fold rotation symmetry. In particular, the curves with $(n_1,n_2)=(1,n)$ possess $(n+1)$-fold rotation symmetry which we use to induce electric polarization by modifying the symmetry of the driven system.}
        \label{RoseCurve}
    \end{center}
\end{figure}

In this paper, we focus on the possibility of Floquet engineering of electric polarization. To this end, we need to control the rotational symmetry of the driven system since rotational symmetry prohibits the emergence of electric polarization. This can be achieved by tailoring the rotational symmetry of the driving electric field. In particular, a two-frequency drive is a suitable platform to engineer the rotational symmetry of the system \cite{Nag19,trevisan21}. 
Specifically, we consider a bicircular light (BCL) defined with the vector potential that reads
\begin{align}
    A(t) = A_\textrm{L} e^{in_1\omega t} + A_\textrm{R} e^{-in_2\omega t+i\theta},
\end{align}
where $A_{\textrm{L}(\textrm{R})}$ is the amplitude of the left (right) circular light and $n_1,n_2$ are the integers associated with the frequencies of the two circular waves.
By tuning relative frequency of the two driving lights, a trajectory of the electric field of the BCL gives so called rose curves, which shows $(n_1+n_2)/\mathrm{gcd}(n_1,n_2)$-fold rotational symmetry for the pair of the integers $(n_1, n_2)$, as illustrated in Fig. \ref{RoseCurve}. Even when the original electronic system possesses a rotational symmetry and exhibits no polarization, application of BCL driving with a rotational symmetry that is incompatible with the original one makes the driven system polar and can generate nonzero electric polarization.

Pursuing the above scenario, we study Floquet engineering of electric polarization with a two-frequency drive. We derive a Floquet Hamiltonian under the driving with BCL and perform high frequency expansion to deduce an effective Hamiltonian that describes the nonequilibrium steady state. 
Applying this method to 1D and 2D tight-binding models, we demonstrate that electric polarization $P$ can be induced by using suitable BCL that invalidates the rotational symmetry of the system. Furthermore, we show that the relative phases of the BCL can control the direction of $P$. We also consider a 2D honeycomb tight-binding model and discuss the possibility of Floquet engineering of electric polarization in materials including hexagonal boron nitride (BN), transition metal dichalcogenide (TMD), and bilayer graphene.

The rest of this paper is organized as follows. In Sec. \ref{sec: formalism}, we introduce our formalism for Floquet engineering of electric polarization with BCLs. 
In Sec. \ref{sec: application}, we apply our method to 1D SSH model, a 2D square lattice model, and a 2D honeycomb lattice model, and demonstrate dynamical control of electric polarization with BCLs.
In Sec. \ref{sec: discussions}, we present a brief discussion.

\section{Formalism \label{sec: formalism}}
In this section, we present our formalism to study electric polarization in the driven system. First, we review Floquet theory which is a time direction analog of Bloch's theorem.
Floquet theory enables us to study nonequilibrium steady states using an effectively static Hamiltonian, i.e.,  Floquet Hamiltonian \cite{andre-RMP,Oka-Kitamura19}. 
The Floquet bands obtained by diagonalizing the Floquet Hamiltonian are periodic in the energy direction as an analog of Brillouin zone in the Bloch's theorem. 
We can construct an effective Hamiltonian for the ``first Brillouin zone'' by a perturbation theory when the driving frequency is high, which is known as a high frequency expansion.
We then use this effective Hamiltonian to compute electric polarization in the driven system using the Berry phase formula \cite{resta-RMP,vanderbilt-kingsmith}.

\subsection{Floquet theory}
First, we introduce Floquet theory.
We consider a quantum system driven by a time-periodic external field with a Hamiltonian satisfying
\begin{align}
    H(t+T)=H(t),
\end{align}
where $T=2\pi/\omega$ is the period of external fields.
Due to the discrete time-translation symmetry of the system, 
the solution of the time-dependent Schr\"{o}dinger equation
$i \dv{t}\ket{\psi(t)}=H(t)\ket{\psi(t)}$ can be taken as an eigenstate of the discrete translation (we set $\hbar=1$).
Namely, the general solution of the time-dependent Schr\"{o}dinger equation can be 
expanded by the Floquet states $\{ \ket{\psi_\alpha (t)} \}$ of the form 
\begin{align}\label{Floquet_Theorem}
    \ket{\psi_\alpha (t)}=e^{-i \epsilon_\alpha t} \ket{u_\alpha (t)},\  \ket{u_\alpha (t+T)}=\ket{u_\alpha (t)},
\end{align}
where $\epsilon_\alpha$ is called Floquet quasi-energy.
Equation (\ref{Floquet_Theorem}) shows that Floquet theory is based on an analog of Bloch's theorem, which is widely known in solid state physics.
Namely, discrete spatial-translation symmetry in Bloch's theorem corresponds to discrete time-translation symmetry in Floquet theory.

We can reinterpret the time-dependent Schr\"odinger equation in the original Hilbert space $\mathcal{H}$ as a static eigenvalue problem in the extended Hilbert space spanned with the original Hilbert space and an additional Floquet index, by expanding the 
time-periodic part $\ket{u_\alpha(t)}$ of the solution of the Schr\"odinger equation as a Fourier series. 
Substituting Floquet states $\ket{\psi_\alpha (t)}$ to 
the time-dependent Schr\"{o}dinger equation and performing some manipulations, we arrive at 
\begin{align}\label{EV_of_EHS}
    \sum_{m\in\mathbb{Z}}
    (H_{n-m}-m\hbar\omega \delta_{m,n} 
    )
    \ket{u_{\alpha}^n}=
    \epsilon_\alpha\ket{u_{\alpha}^n},
\end{align}
where $H_n = \frac{1}{T}\int_0^T \dd t e^{in\omega t}H(t)$
 and $\ket{u_{\alpha}^n}=\frac{1}{T}
    \int_0^T \dd t e^{in\omega t}
    \ket{u_{\alpha}(t)}$ are the $n$-th Fourier coefficients of $H(t)$ and $\ket{u_\alpha(t)}$, respectively.
Equation (\ref{EV_of_EHS}) is an eigenvalue equation in the extended Hilbert space~\cite{Shirley1965,Sambe1973}.

\subsection{High frequency expansion}


Because $\ket{u_{\alpha^\prime} (t)}=\ket{u_\alpha (t)}e^{i m\omega t}$ also satisfies $\ket{u_{\alpha^\prime} (t)}=\ket{u_{\alpha^\prime} (t+T)}$,
there are multiple representations for the same Floquet state $\ket{\psi_\alpha(t)}$ in the extended Hilbert space.
This implies that the eigenvalue problem (\ref{EV_of_EHS}) includes many redundant solutions with $\epsilon_{\alpha^\prime}=\epsilon_{\alpha}+m\omega$.
This redundancy can be removed by block-diagonalizing the matrix in Eq.~(\ref{EV_of_EHS}),
after which the matrix has the form $(H_\text{eff}-m\hbar\omega)\delta_{m,n}$ with the block matrix $H_\text{eff}$ being an $m$-independent operator on $\mathcal{H}$. 

One approach to perform block-diagonalization and remove such redundancy of a Floquet Hamiltonian is known as van Vleck's degenerate perturbation theory, which is applicable when  the driving frequency of the external fields $\omega$ is much larger than the energy scale of the system \cite{Eckardt2015,Mikami16}.
In such a situation, application of van Vleck's degenerate perturbation theory leads to the effective Hamiltonian $H_{\textrm{eff}}$ in the form,
\begin{align}
    H_{\textrm{eff}}^{\textrm{vV}}=
    H_0+
    \sum_{m\neq 0}\left(
    \frac{[H_{-m},H_m]}{2m\omega}+
    \frac{[[H_{-m},H_0],H_m]}{2m^2\omega^2}+
    \sum_{n\neq0,m}\frac{[[H_{-m},H_{m-n}],H_n]}{3mn\omega^2}
    \right)
    +\mathcal{O}(\omega^{-3})\label{eq:HFE}
\end{align}
by taking the $-m\hbar\omega\delta_{m,n}$ term as the unperturbed part and 
treating the $H_{n-m}$ term as a perturbation.

\subsection{Berry phase formula for electric polarization}
Let us briefly explain Berry phase theory of electric polarization \cite{resta-RMP,vanderbilt-kingsmith}.
From elementary adiabatic perturbation theory, electric current from the $n$-th band is obtained as 
\begin{align}
    \vb*{j}_n(\lambda)=\dv{\vb*{P}_n(\lambda)}{t}=
    \frac{i e \Dot{\lambda}}{(2\pi)^d m}
    \sum_{l\neq n}
    \int \dd^d \vb*{k} 
    \frac{\bra{u_{n\vb*{k}}}\vb*{p}\ket{u_{l\vb*{k}}} \bra{u_{l\vb*{k}}} \partial_\lambda \ket{u_{n\vb*{k}}}}{E_{n\vb*{k}}-E_{l\vb*{k}}}+ \textrm{c.c.},
\end{align}
where $\lambda\in [0,1]$ denotes the adiabatic parameter and $\ket{u_{n\vb*{k}}}$ is the $n$-th Bloch's state.
After integration with respect to $\lambda$, we get Berry phase formula:
\begin{align}
    \vb*{P}(\lambda)=
    \frac{-ie}{(2\pi)^d}\sum_{n:\textrm{occ.}}
    \int \dd^d \vb*{k} \bra{u_{n\vb*{k}}}\grad_{\vb*{k}}\ket{u_{n\vb*{k}}},
\end{align}
or in the gauge invariant form
\begin{align}\label{gauge_invariant_P}
    \Delta \vb*{P} = \vb*{P}(1)-\vb*{P}(0)=
     \frac{ie}{(2\pi)^d}\sum_{n:\textrm{occ.}}
    \int \dd^d \vb*{k} \int_0^1 \dd \lambda  \braket{\grad_{\vb*{k}} u_{n\vb*{k}}}{\partial_\lambda u_{n\vb*{k}}}
    +\textrm{c.c.}
\end{align}
The integrand function of Eq. (\ref{gauge_invariant_P}) is the Berry curvature in the parameter space $(\vb*{k},\lambda)$.
Hence, electric polarization is expressed in the form of the Berry phase.

\subsection{Two frequency drive}
In systems with certain symmetries, polarization often vanishes. If a system has spatial inversion symmetry $\mathcal{I}$, the inversion operation will result in $P\to -P$, so that $P=0$ is required. Similarly, if a system has $n$-fold symmetry $C_n$, $P= 0$ is also required. Bicircular light (BCL) used in this study is convenient as an external field to break such symmetries.
By choosing an appropriate frequency ratio, BCL can break not only $\mathcal{I}$ symmetry but also $C_n$ symmetry.

Bicircular light consists of two circular light (CL) waves with different frequencies and opposite chirality. BCL can be expressed in the form of a gauge field $A(t)=A_x(t)+ i A_y(t)$ as
\begin{align}
    A(t) = A_\textrm{L} e^{in_1\omega t} + A_\textrm{R} e^{-in_2\omega t+i\theta},
\end{align}
where $A_{\textrm{L}(\textrm{R})}$ is the amplitude of the left (right) CL and $n_1,n_2$ are the integers representing the frequencies of the two CL waves. As shown in Fig.~\ref{RoseCurve}, BCL waves draw various rose curves with $n_1$ and $n_2$. The number of leaves on the rose curve is determined by the integer ratio of $n_1$ to $n_2$.
In particular, we focus on the case of $(n_1,n_2)=(1,n)$ in this paper, where the gauge field is given by
\begin{align}
    A(t) = A_\textrm{L} e^{i\omega t} + A_\textrm{R} e^{-in\omega t+i\theta}.
\end{align}
This gauge field has $(n+1)$-fold symmetry, so if a system has $C_m$ symmetry, tuning the frequency ratio $n$ so that $n+1$ is coprime to $m$ can break $C_m$ symmetry and can induce electric polarization.
The parameter $\theta$ is the phase difference between two CL waves and controlling it can rotate the rose pattern drawn by BCL.
Rotation of the rose pattern is expected to cause rotation of the electric polarization direction.

\section{Floquet engineering of electric polarization \label{sec: application}}
In this section, we demonstrate dynamical control of electric polarization by using two-frequency drive. 
First, we present a result for 1D Su-Shrieffer-Heeger (SSH) model and a 2D square lattice model. 
We then discuss the possibility of Floquet engineering of electric polarization in honeycomb lattice model with staggered potential which is relevant to $C_3$ symmetric 2D materials such as hexagonal boron-nitride (BN), transition metal dichalcogenides (TMD) and bilayer graphene.

\subsection{1D SSH Model \label{subsec: 1d model}}
\begin{figure}[t]
    \begin{center}
        \includegraphics[width=\linewidth]{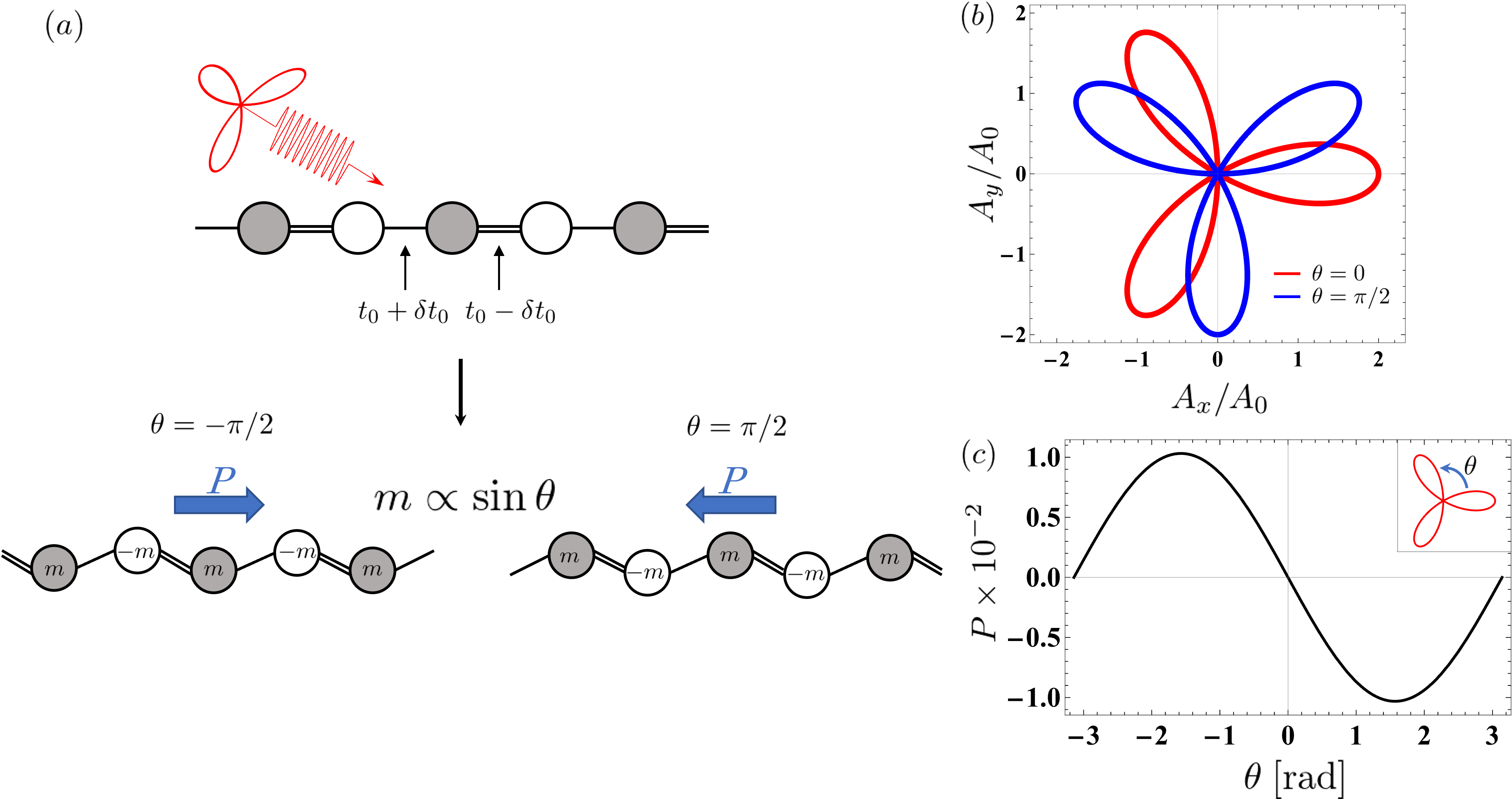}
        \caption{Floquet engineering of SSH model with BCL. (a) A schematic picture of SSH model. Hopping amplitude shows alternation with $t_0\pm \delta t_0$. Application of 3-fold BCL induces electric polarization. (b) 3-fold rotation symmetric electric field of the BCL for $\theta=0,\pi/2$.
        (c) $\theta$-dependence of the electric polarization $P$ under the drive.
        We used the parameters, $t_0=1$, $\delta t_0=0.5$, $A_0=0.5$, and $\omega=2$.}
        \label{fig2}
    \end{center}
\end{figure}

Su-Schrieffer-Heeger (SSH) model is a famous model for polyethylene, and consists of 1D tight-binding model with bond alternation \cite{su79}.
The SSH model has mirror symmetry with respect to the bond center and hence supports no polarization.
We show that subjecting the SSH model to two frequency drive with BCL effectively realize Rice-Mele model \cite{rice-mele}, as illustrated in Fig.~\ref{fig2}(a).
Rice-Mele model is a representative model for 1D ferroelectrics which breaks inversion symmetry and supports nonzero electric polarization.

The SSH model is described by the two-band Hamiltonian which is given by
\begin{align}
    \mathcal{H}=\sum_{j=1}^N [(t_0+\delta t_0) c_{j,A}^{\dagger}c_{j,B}+ (t_0-\delta t_0) c_{j,A}^{\dagger}c_{j-1,B} ] + \textrm{h.c.},
\end{align}
where $c_{j,X}~(c_{j,X}^\dagger)$ is the annihilation (creation) operator for site $X$ of the $j$-th unit cell and $t_0$ is the hopping parameter.
Fourier transformation of creation and annihilation operators yields Bloch Hamiltonian (we set lattice constant to be $a=2$):
\begin{align}
    H(k)=2t_0 \cos k \sigma_x -2\delta t_0 \sin k\sigma_y,
\end{align}
where $\sigma_i$ $(i=x,y,z)$ are Pauli matrices.
$H(k)$ has inversion symmetry $\mathcal{I}$, i.e., $\sigma_x H(k) \sigma_x=H(-k)$, so that the electric polarization does not appear%
\footnote{More precisely, the electric polarization $P$ satisfies $P=-P$ mod $a$ when the inversion symmetry is preserves. This condition is satisfied either by vanishing polarization $P=0$ or by the half of the lattice constant $P=a/2$. In the latter case, the polarization can be nullified  by changing the choice of the unit cell.}.
We can break $\mathcal{I}$ symmetry by applying a $C_{3}$ symmetric electric field, which can be realized by 3-fold BCL:
\begin{align}
    A(t)=\Re [A_0 e^{i\omega t} + A_0 e^{-2i \omega t +i \theta} ]
\end{align}
Figure \ref{fig2}(b) illustrates the rose patterns drawn by this gauge field $A(t)$ for $\theta=0,\pi/2$.
Application of the 3-fold BCL to the system leads to the time-dependent Hamiltonian (we set $e=1$)
\begin{align}\label{FloquetBloch_1}
    H(k+A(t))=2t_0 \cos [k+A(t)] \sigma_x -2\delta t_0 \sin [k+A(t)]\sigma_y.
\end{align}
From Jacobi–Anger expansion
$
    e^{iz\cos \phi}=
    \sum_{n\in \mathbb{Z}} i^n \mathcal{J}_n(z)e^{in\phi},
$
($\mathcal{J}_n$: Bessel functions of the first kind), the time-dependent Hamiltonian (\ref{FloquetBloch_1}) can be written as
\begin{align}\label{Ham_1D_Bessel}
\begin{split}
    H(t)=2t_0 \Re &\left[
    \sum_{n,m}i^{n+m}e^{i(n+2m)\omega t}e^{ik}e^{-im\theta}\mathcal{J}_n(A_0)\mathcal{J}_m(A_0)
    \right]\sigma_x\\
    -2\delta t_0 \Im &\left[
    \sum_{n,m}i^{n+m}e^{i(n+2m)\omega t}e^{ik}e^{-im\theta}\mathcal{J}_n(A_0)\mathcal{J}_m(A_0)
    \right]\sigma_y.
\end{split}
\end{align}
This leads to the Fourier coefficients $H_m$ for any BCL amplitude $A_0$. Here, assuming a small amplitude and truncating the higher order Bessel functions, we expand Eq. (\ref{Ham_1D_Bessel}) up to the second order of $A_0$, which yields
\begin{gather}
    H_0=(2-A_0^2)
    \biggl[
    t_0\cos k \sigma_x 
-\delta t_0 \sin k \sigma_y 
\biggr],
\\
H_{\pm 1}=
    t_0 \biggl[
    - A_0 \sin k - \frac{1}{2} A_0^2 \cos k  e^{\pm i\theta}    
    \biggr]
    \sigma_x 
    -\delta t_0 
    \biggl[
     A_0 \cos k - \frac{1}{2} A_0^2\sin k  e^{\pm i\theta}    
    \biggr]
    \sigma_y ,
\\
H_{\pm 2}=
    t_0 \biggl[
    - A_0 \sin k e^{\pm i\theta} - \frac{1}{4} A_0^2 \cos k     
    \biggr]
    \sigma_x 
    -\delta t_0 
    \biggl[
     A_0 \cos k e^{\pm i\theta} - \frac{1}{4} A_0^2 \sin k 
    \biggr]
    \sigma_y,
\\
    H_{\pm 3}=
   -\frac{1}{2} t_0 A_0^2 \cos k e^{\pm i\theta}\sigma_x
   +\frac{1}{2} \delta t_0 t_0 A_0^2 \sin k e^{\pm i\theta}\sigma_y,
\\
    H_{\pm 4}=
   -\frac{1}{4} t_0 A_0^2 \cos k e^{\pm 2i\theta}\sigma_x
   +\frac{1}{4} \delta t_0 t_0 A_0^2 \sin k e^{\pm 2i\theta}\sigma_y.
\end{gather}
By computing commutators $[H_{-m},H_m]$, the effective Hamiltonian [Eq.~(\ref{eq:HFE})] up to the first order of $1/\omega$ is obtained as
\begin{align}\label{1D_SSH_Heff}
   H_{\textrm{eff}}=H_0+
\frac{3}{2\omega}t_0\delta t_0  A_0^3 \sin\theta \sigma_z
+\mathcal{O}(\omega^{-2},A_0^4).
\end{align}
This shows that a mass term ($\propto \sigma_z$) appears and Rice-Mele model is realized effectively by the two frequency drive.

Since a mass gap opens in the effective Hamiltonian, we can define electric polarization of the lower band by using the Berry phase formula as 
$P(\theta)=-i\int_{\textrm{BZ}} \frac{\dd k}{2\pi} 
    \bra{u_{-}(k,\theta)}\partial_k \ket{u_{-}(k,\theta)}$,
where $\ket{u_- (k,\theta)}$ is the wave function of the lower band.
This gives the electric polarization of the driven system if the lower band of the effective Hamiltonian is fully occupied in the steady state under the driving%
\footnote{In the nonequilibrium steady state under the driving, the electron distribution generally depends on the relaxation process. For example, with small dissipation, the energy of the system increases due to the energy gain from the drive and its temperature may approach infinity~\cite{DAlessio2014,Lazarides2014}.
In order that the lower energy states are (fully) occupied in the steady state, we need efficient relaxation processes such as phonon scattering.}.
In a two-level system with a Hamiltonian $H=\vb*{R}\vdot \vb*{\sigma}$, Berry connection of the lower band is written as 
\begin{align}
    a_-(k)
    =\frac{1}{2}(1+\cos \xi)\pdv{\eta}{k},
\end{align}
where $\vb*{R}=R(
    \sin \xi \cos \eta, \sin \xi \sin \eta,    \cos \xi)$.
In this model, $\vb*{R}=((2-A_0^2)t_0\cos k,-(2-A_0^2)\delta t_0\sin k,\frac{3}{2\omega}t_0\delta t_0 A_0^3 \sin \theta)$, which leads to
\begin{align}\label{Eq:P_1D}
    P(\theta)=-\frac{3}{8\pi}\frac{1}{\omega}t_0\delta t_0 A_0^3 \sin\theta\int_{-\pi/2}^{\pi/2} \dd k \frac{1}{|\vb*{R}|}
    \frac{\sec^2k}{1+(\delta t_0/t_0)^2 \tan^2k}.
\end{align}
Figure~\ref{fig2}(c) plots the electric polarization $P$ as a function of $\theta$, and as Eq.~(\ref{Eq:P_1D}) indicates, if $A_0\ll 1$, then $P$ is proportional to $-\sin\theta$.
Since $P$ is expressed as an odd function of $\theta$,
we can dynamically control the sign of the electric polarization by tuning the sign of $\theta$.

\subsection{2D square lattice model}\label{subsec: 2d model}
\begin{figure}[t]
    \begin{center}
        \includegraphics[width=\linewidth]{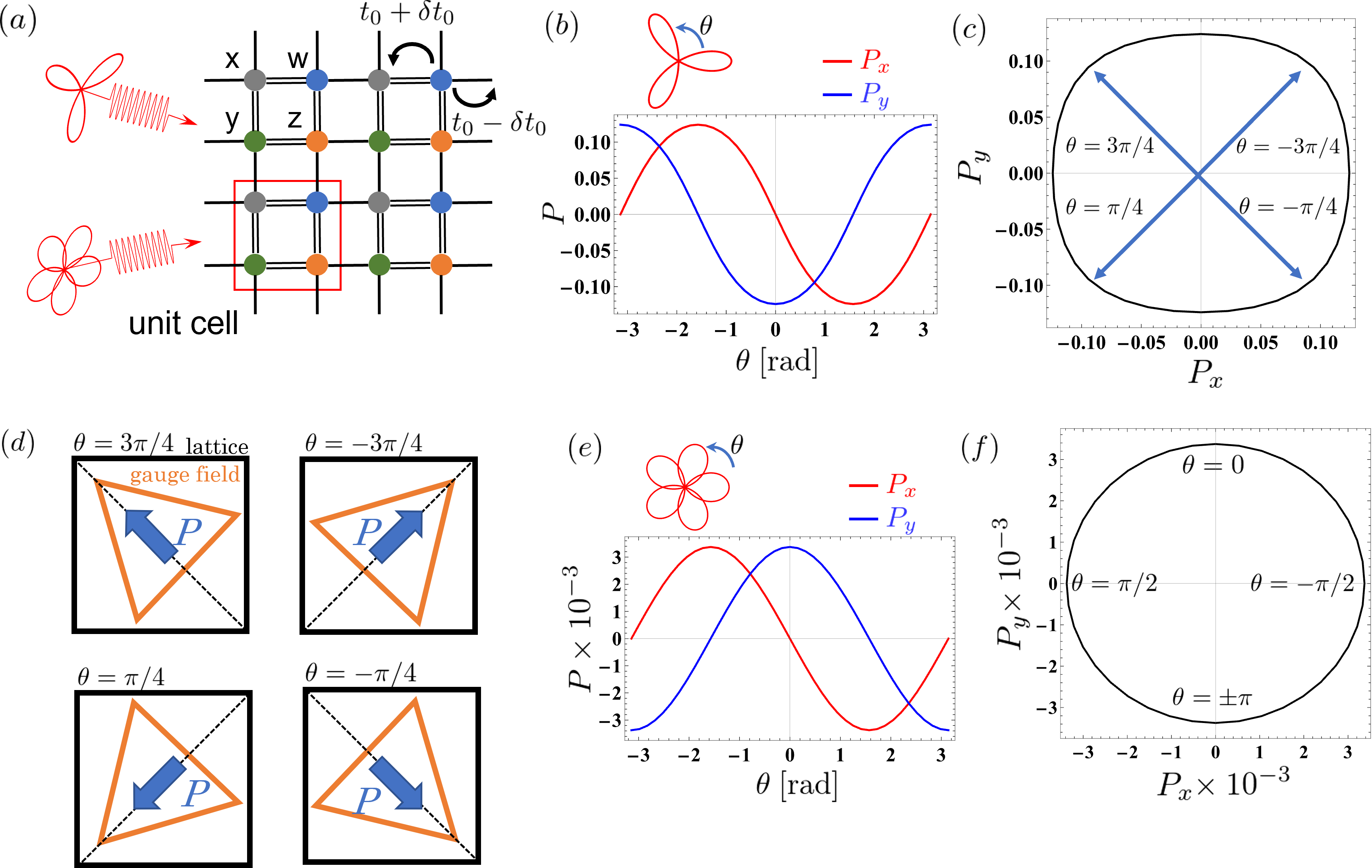}
        \caption{Floquet engineering of electric polarization in the two dimensional square lattice with BCL.
        (a) Schematic picture of the 2D toy model. Tight binding model on the 2D square lattice is subjected to $C_3$ and $C_5$ symmetric BCL to break rotational symmetry and induce electric polarization. (b,c) Electric polarization induced by 3-fold BCL. Polarization direction can be controlled by the relative phase $\theta$ of the BCL. (d) Schematic picture indicating the direction of polarization that can be deduced from relative patterns of the $C_4$ symmetric lattice and the $C_3$ symmetric gauge fields. 
        (e,f) Electric polarization induced by 5-fold BCL. 
        We used the parameters, $t_0=1$, $\delta t_0=0.5$, $A_0=1$, and $\omega=2$.}
        \label{fig3}
    \end{center}
\end{figure}
Next, we consider a 2D toy model defined on a square lattice as illustrated in Fig~\ref{fig3}(a). 
The unit cell consists of 4 sites forming a square, labelled X, Y, Z, and W in Fig.~\ref{fig3}(a).
The intra-unit-cell hopping amplitude is $t_0+\delta t_0$ and the inter-unit-cell hopping amplitude is $t_0-\delta t_0$.
This model can be regarded as an extension of 1D SSH model.
The Hamiltonian is given by $\mathcal{H}=\sum_{\vb*{k}} \vb*{c_k}^\dagger H(\vb*{k})\vb*{c_k}$ with annihilation operators $\vb*{c_k}=(c_{X\vb*{k}},c_{Y\vb*{k}},c_{Z\vb*{k}},c_{W\vb*{k}})^\top$ and the Bloch Hamiltonian
\begin{align}
\begin{split}
    H(\vb*{k})\!=\!2 
\mqty(
    0 & t_0\!\cos\! k_y \!-\! i \delta t_0\! \sin\! k_y & 0 & t_0\!\cos\! k_x \!+\! i \delta t_0\!\sin\! k_x \\
    t_0\!\cos\! k_y \!+\! i \delta t_0\!\sin\! k_y & 0 & t_0\!\cos\! k_x \!+\! i \delta t_0\!\sin\! k_x & 0   \\
    0 & t_0\!\cos\! k_x \!-\! i \delta t_0\!\sin\! k_x & 0 & t_0\!\cos\! k_y \!+\! i \delta t_0\!\sin\! k_y \\
    t_0\!\cos\! k_x \!-\! i \delta t_0\!\sin\! k_x & 0 & t_0\!\cos\! k_y \!-\! i \delta t_0\!\sin\! k_y & 0
),
\end{split}
\end{align}
where we set the lattice constant to be $a=2$.
This Bloch Hamiltonian $H(k)$ has a 4-fold rotational symmetry $C_4$, i.e.,
\begin{align}\label{Eq:symmetry_system}
    U_4^\dagger H(\eta) U_4&=H(\eta+\frac{\pi}{2}), & 
    U_4&=\mqty(
    0&1&0&0\\
    0&0&1&0\\
    0&0&0&1\\
    1&0&0&0
    ),
\end{align}
where $\eta=\arg (k_x+ik_y)$ and $U_4$ is the matrix representing the $C_4$ symmetry.
In a similar manner as in Sec. \ref{subsec: 1d model}, this symmetry can be broken by applying a 3-fold BCL, 
$A_x+iA_y=A_0(e^{i\omega t} + e^{-2i \omega t +i \theta})$, because of $\mathrm{gcd}(4,3)=1$.
The effective Hamiltonian under the driving is obtained by a similar calculation as in the previous section (for details, see Appendix \ref{app: 2d model}):
\begin{align}\label{Eq:EH_2D_C3}
    H_{\text{eff}}
    =H_0 +
\frac{3}{2\omega}t_0\delta t_0  A_0^3
\biggl[
    \sin \theta (\sigma_z \otimes  I)
    - \cos \theta (\sigma_z \otimes  \sigma_z)
\biggr] 
+\mathcal{O} (\omega^{-2} , A_0^4),
\end{align}
where $H_0=\frac{2-A_0^2}{2}H(k)$ and $\otimes$ denotes tensor product defined as
\begin{align}
    A\otimes B=\mqty(
    a_{11} B & a_{12} B\\
    a_{21} B & a_{22} B
    )
\end{align}
with $2\cross 2$ matrices $A=(a_{ij})$ and $B$.
This effective Hamiltonian includes the terms $\propto \sigma_z$ and can be regarded as a 2D extension of the 1D Rice Mele model that appeared in the 1D case (see Eq. (\ref{1D_SSH_Heff}) for comparison).
Therefore, application of a 3-fold BCL can induce electric polarization in the square lattice model. 

Again assuming that only the lowest band is occupied (with an efficient relaxation), we can obtain the electric polarization from the Berry phase formula, as shown in Fig.~\ref{fig3}(b,c).
We can control the direction of the polarization $(P_x, P_y)$ with the relative phase of two frequency drive $\theta$. 
As in Fig.~\ref{fig3}(b), the fact that $P_x$ $(P_y)$ is an odd (even) function of $\theta$ can be confirmed from the symmetry argument. The effective Hamiltonian Eq.~(\ref{Eq:EH_2D_C3}) satisfies 
\begin{align}
    &U^\dagger H_{\text{eff}}(-k_x,k_y,-\theta)U
    =H_{\text{eff}}(k_x,k_y,\theta),
    &U=\sigma_x \otimes \sigma_x,
\end{align}
where the unitary transformation $U$ is the mirror operation on $y$-axis.
Thus, it is derived that $P_x(-\theta)=-P_x(\theta)$ and $P_y(-\theta)=P_y(\theta)$.
Eq.~(\ref{Eq:EH_2D_C3}) indicates that  $P_x(\theta)$ and $P_y(\theta)$ are proportional to $-\sin\theta$ and $-\cos\theta$, respectively, when $A_0 \ll 1$, as in 1D SSH model.
However, when $A_0$ becomes larger, the $P_x$-$P_y$ curve shows anisotropy and looks like a rounded square as shown in Fig.~\ref{fig3}(c), where
$P=\sqrt{P_x^2+P_y^2}$ is peaked at $\theta=\pm\pi/4,\pm 3\pi/4$.
Figure~\ref{fig3}(d) schematically illustrates the reason why $P_x$-$P_y$ curve shows a $C_4$ symmetric pattern.
In Fig.~\ref{fig3}(d), the underlying $C_4$ symmetry of the square lattice is depicted as squares and the gauge field of BCL with the $C_3$ symmetry as equilateral triangles.
(We note that the direction of the triangle is different from the rose curve pattern actually drawn by the gauge field.)
When a bisection of the triangle coincides with the diagonal of the square lattice, which happens at $\theta=\pm\pi/4,\pm 3\pi/4$, the diagonal behaves as a mirror plane for the overall structure, which constrains the direction of electric polarization in the diagonal direction.
As the direction of polarization rotates by $\pi/2$, the $P_x$-$P_y$ curve shows a rounded square pattern reflecting the $C_4$ symmetry of the square lattice.

From a more precise symmetry argument, we can prove that the electric polarization reflects the symmetry of the square lattice as follows.
We find that the time-dependent Bloch Hamiltonian $H(k_x,k_y,t,\theta)$ 
satisfies the relationship, 
\begin{align}
    U_4^\dagger 
    H(k_y,-k_x,t-\pi/2\omega,\theta+\pi/2)
    U_4
    =H(k_x,k_y,t,\theta),
\end{align}
where $U_4$ defined in Eq.~(\ref{Eq:symmetry_system}) is an operator that rotates the system by $\pi/2$.
(For details of the derivation and its generalization, see Appendix \ref{app: symmetery polarization}.)
This equation indicates that the shift of the relative phase $\theta \to \theta +\pi/2$ and the time translation $t \to t-\pi/2\omega$ results in $(-\pi/2)$-rotation of the system characterized by $(k_x,k_y) \to (k_y,-k_x)$. This immediately leads to the $C_4$ symmetry of the electric polarization $P$ which can be expressed as
\begin{align}\label{Eq:C4_P}
    &P_x(\theta+\pi/2)=P_y(\theta),
    &P_y(\theta+\pi/2)=-P_x(\theta),
\end{align}
as observed in Fig.~\ref{fig3}(c).
Furthermore, this dynamical symmetry is associated with the unitary operator $U_4$ that describes the $C_4$ symmetry of the square lattice model in the equilibrium [Eq.~\eqref{Eq:symmetry_system}]. This implies that the symmetry of the electric polarization with respect the phase $\theta$ reflects the symmetry of the underlying lattice of the system, which generally holds as we see for the case of $C_3$ symmetric lattice model in the next subsection.

$C_5$ symmetric BCL 
$A_x+iA_y=A_0(e^{i\omega t} + e^{-4i \omega t +i \theta})$
can also break $C_4$ symmetry of the 2D model, because of $\mathrm{gcd}(4,5)=1$.
By using up to the third order of $A_0$ in the Fourier coefficients, we can obtain another effective Hamiltonian including mass terms,
\begin{align}
    H_{\text{eff}}
    =H_0 +
\frac{5}{144\omega}t_0\delta t_0 A_0^5
\biggl[
    \sin \theta (\sigma_z \otimes  I)
    + \cos \theta (\sigma_z \otimes  \sigma_z)
\biggr] 
+\mathcal{O} (\omega^{-2} , A_0^6).
\end{align}
This is almost the same as the effective Hamiltonian in the case of $C_3$ symmetric BCL driving [Eq. (\ref{Eq:EH_2D_C3})].
Figure~\ref{fig3}(e,f) shows the results of the electric polarization when $C_5$ BCL is driven on the 2D square lattice.
Because of the need for higher order of Fourier coefficients $H_m$ and the BCL amplitude $A_0$ to produce the mass terms, the mass terms are smaller than in the case of $C_3$ symmetric BCL for the same $A_0$. Thus the electric polarization is also smaller and anisotropy of the electric polarization almost disappears.
Note that the rotation direction of the electric polarization is opposite to that of the $C_3$ symmetric BCL driven case due to the difference in the shape of the rose patterns drawn by BCLs. Appendix \ref{app: symmetery polarization} gives a more description of the relationship between the shape of the rose curve and the rotation direction of the electric polarization.

\subsection{Honeycomb lattice model}

\begin{figure}[t]
    \begin{center}
        \includegraphics[width=14cm]{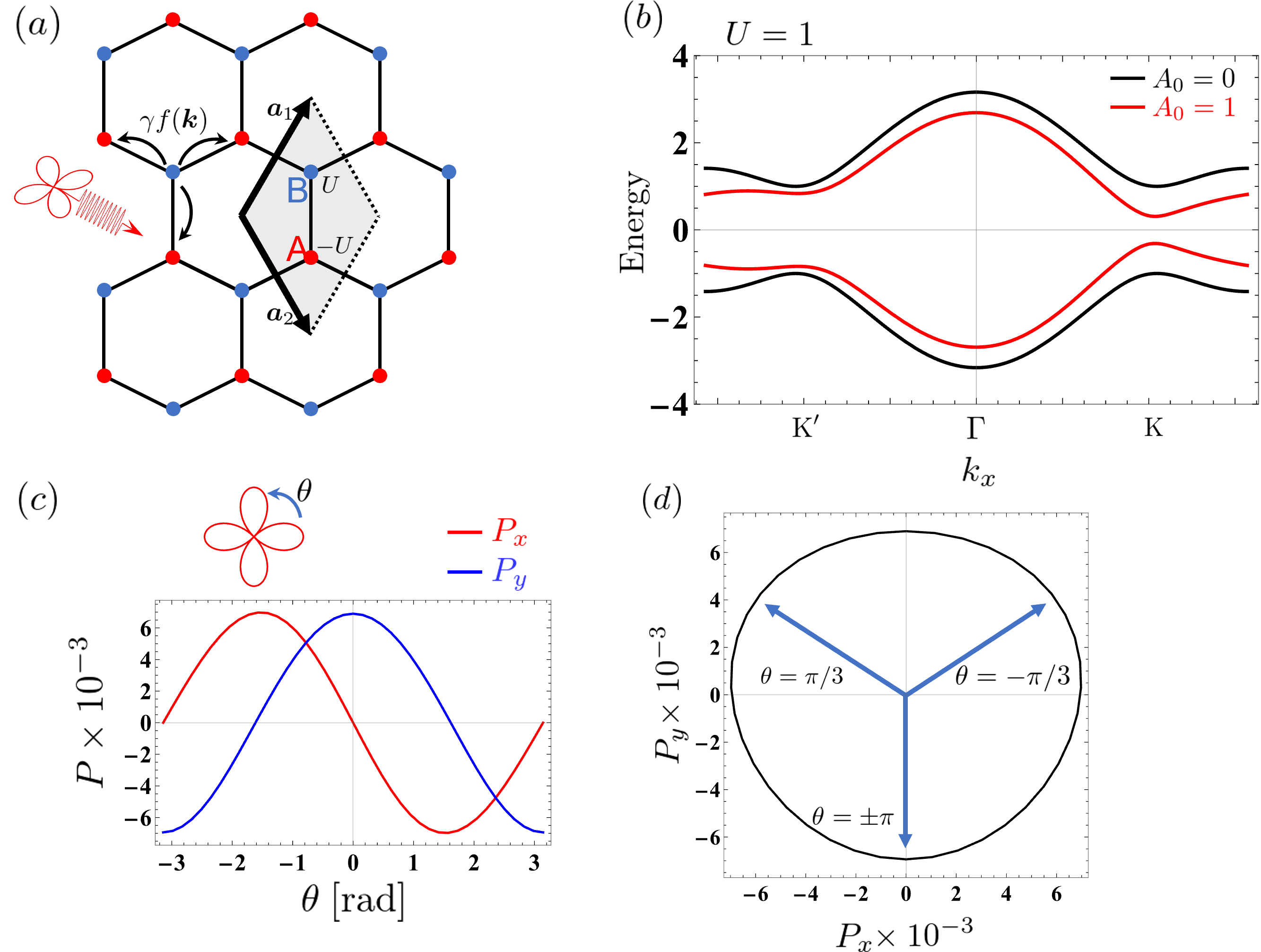}
        \caption{Floquet engineering of electric polarization in the two dimensional honeycomb lattice model with $C_4$ symmetric BCL. 
        (a) Tight binding model on the honeycomb lattice. Staggered potential $U$ is introduced for two sublattices labeled by A and B. 
        (b) The band structure of the honeycomb lattice along $k_x$ axis, without the driving ($A_0=0$, black) and with the driving ($A_0=1$, red). $\Gamma$ and K (K$'$) indicate the center and the corners of the Brillouin Zone, respectively.
        Since K and K$'$ points are not equivalent with the driving, the band gaps are different at the two points, resulting in the asymmetric band structure.
        (c) Electric polarization $P$ as a function of the relative phase $\theta$ of BCL.
        (d) Trajectory of induced polarization in the $P_x$-$P_y$ plane. The trajectory shows directional anisotropy reflecting $C_3$ symmetry of the honeycomb lattice.
        We used the parameters, $a=1$, $U=1$, $\gamma =1$, $A_0=1$, and $\omega=2$.
        }
        \label{fig4}
    \end{center}
\end{figure}

Finally, we apply our method to Floquet-engineer electric polarization to the honeycomb lattice. 
Figure ~\ref{fig4}(a) shows the structure of the honeycomb lattice. 
Primitive lattice vectors $\vb*{a}_1$ and $\vb*{a}_2$ are defined as
\begin{align}
\vb*{a}_1 =\left(\frac{a}{2},\frac{\sqrt{3}a}{2}\right),\ 
\vb*{a}_2 =\left(\frac{a}{2},\frac{-\sqrt{3}a}{2}\right),
\end{align}
where $a$ is the lattice constant.
Each unit cell, which is grayed out in Fig.~\ref{fig4}(a), contains two sites, labeled A and B.
We further introduce a potential difference $2U$ between A site and B site.
The tight binding Hamiltonian of our honeycomb lattice model is written as
\begin{align}
    H(\vb*{k})=\mqty(
    U & \gamma f(\vb*{k}) \\
    \gamma f^*(\vb*{k}) & -U
    ),
    \label{eq: H honeycomb}
\end{align}
where $\gamma$ is a hopping parameter and $f(\vb*{k})$ represents the hopping from site A to the three nearest-neighbor B sites and can be written as follows using the relative distance to the nearest-neighbor sites $\vb*{\delta}_1=(0,a/\sqrt{3}),~\vb*{\delta}_2=(a/2,-a/2\sqrt{3}),~\vb*{\delta}_3=(-a/2,-a/2\sqrt{3})$:
\begin{align}
    f(\vb*{k})=\sum_{i=1}^3 e^{i\vb*{k}\vdot \vb*{\delta}_i}
    =e^{ik_ya/\sqrt{3}}+2e^{-ik_ya/2\sqrt{3}}\cos(k_xa/2).
\end{align}

Since A and B sites are not equivalent due to the on-site potential difference $U$, this model does not have $C_6$ symmetry but $C_3$ symmetry.
On the basis of the same idea as above, it is expected that the application of BCL with $C_4$ symmetry to this system can break $C_3$ symmetry of the honeycomb lattice and induce the electric polarization (due to $\mathrm{gcd}(3,4)=1$).
Thus we consider application of the BCL with $C_4$ symmetry, $A(t)=A_0(e^{i\omega t} + e^{-3i \omega t +i \theta})$,
to the Hamiltonian Eq.~\eqref{eq: H honeycomb} with the minimal coupling $k\to k+A(t)$.
In this case, we compute the effective Hamiltonian with high frequency expansion by keeping the terms up to the second order with respect to $1/\omega$ in Eq. (\ref{eq:HFE}).
Figure~\ref{fig4}(b) shows the band structure of this honeycomb model with and without $C_4$ symmetric BCL drive at $k_y=0$. Under the $C_4$ symmetric BCL drive, the band gaps are different at K and K$'$ points [$\vb*{k}=(4\pi/3a,0),(-4\pi/3a,0)$] because the electronic states at the K and K$'$ points couple to the BCL differently depending on their chiralities.
This asymmetry in the band structure and the associated wave functions are the origins for the nonzero electric polarization under the driving.

Figure~\ref{fig4}(c,d) shows the result of the calculation of the electric polarization obtained from the effective Hamiltonian as described above.
For $A_0 \ll 1$, $P_x(\theta)$ and $P_y(\theta)$ are almost proportional to $\sin \theta$ and $-\cos \theta$, respectively, 
which clearly indicates that one can control the direction of the induced polarization by the relative phase $\theta$ of BCL.
When $A_0$ becomes larger, the $P_x$-$P_y$ curve shows a $C_3$ symmetric pattern as shown in Fig.~\ref{fig4}(d), which reflects the underlying $C_3$ symmetry of the lattice structure.
The magnitude of the polarization $P$ shows maximum at $\theta=\pm \pi/3, \pm \pi$. 
Since the results above essentially rely on the $C_3$ symmetry of the lattice structure, these results suggest that such Floquet engineering of electric polarization is possible in general two-dimensional materials with $C_3$ symmetry, which include hexagonal boron nitride (BN), transition metal dichalcogenides such as MoS$_2$, and bilayer graphene with layer asymmetry.

\section{Discussions \label{sec: discussions}}

\begin{figure}[t]
    \begin{center}
        \includegraphics[width=\linewidth]{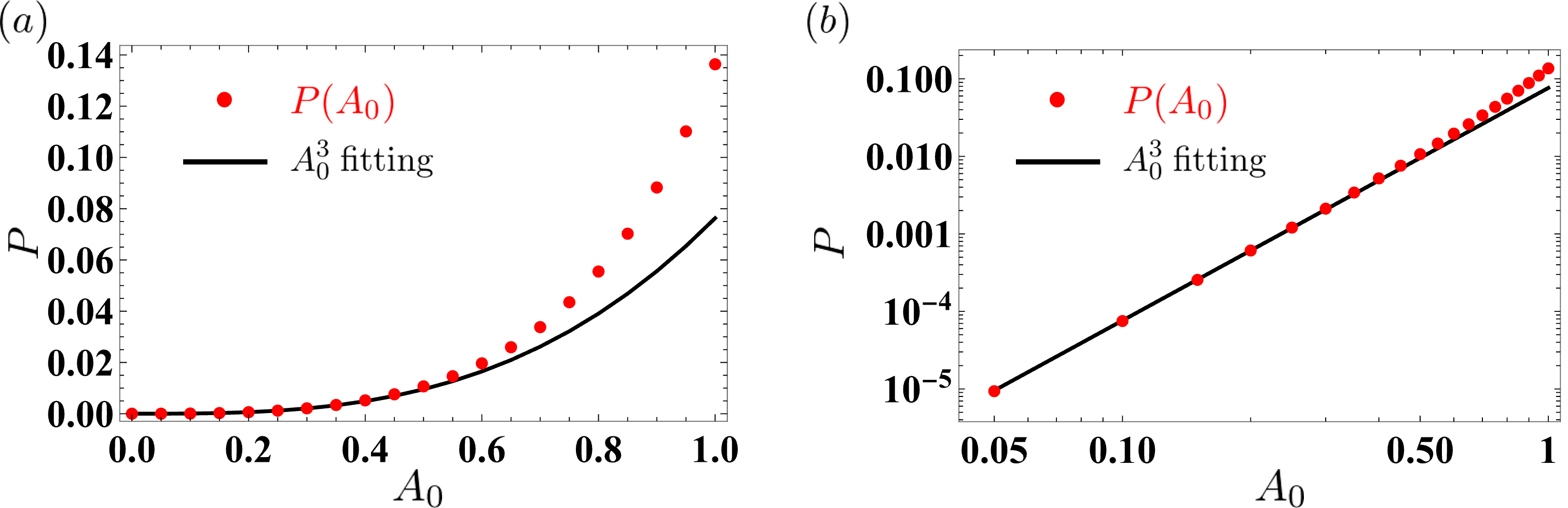}
        \caption{
        Electric field dependence of the induced polarization for the square lattice model.
        (a) Polarization $P$ as a function of the gauge field $A$ (red dots). Low field region is well fitted by $P\propto A^3$ (black curve) as can be seen from Equation (\ref{Eq:P_1D}).
        (b) Log-log plot of $P$ as a function of $A$.
        We used the parameters, $t_0=1$, $\delta t_0=0.5$, and $\omega=2$.
        }
        \label{fig5}
    \end{center}
\end{figure}

We give an order estimate of electric polarization induced by application of BCL. First the strength of the vector potential $A$ is estimated as follows. 
So far the vector potential $A$ was treated as a dimensionless parameter which is given with dimensionful parameters as $A=eEa/\hbar\omega$ with the electric field $E$, the lattice constant $a$, and the photon energy $\hbar \omega$.
For the typical values, $E=10\mathrm{MV}/\mathrm{cm}$, $a=4$\AA~ and the photon energy $\hbar \omega= 2$eV, we obtain
the dimensionless parameter $A=0.2$.
We plot the polarization $P$ as a function of $A$ in Fig.~\ref{fig5} for the square lattice model presented in Sec~\ref{subsec: 2d model}. 
From Fig.~\ref{fig5}, we find that $A=0.2$ can induce the electric polarization of $P_\mathrm{2d}=4 \times 10^{-4}$ for the two-dimensional system.
In the two dimensions, $P_\mathrm{2d}$ measures the excess charge at the edge of the sample for a unit length, where its unit is [$e/a$].
If we assume the two dimensional system is stacked along the $z$ direction with the period of $a$, the resulting electric polarization $P_\mathrm{3d}$ is given by $P_\mathrm{3d}=P_\mathrm{2d}/a$ with the unit [$e/a^2$], which can be compared with polarization of conventional ferroelectric materials. 
With the above 3D extension, $P_\mathrm{2d}=4 \times 10^{-4} e/a$ induced by BCL in the square lattice model amounts to $P_\mathrm{3d}=4 \times 10^{-4}$C/m$^2$.
For comparison, the polarization in typical ferroelectric material BaTiO$_3$ is $0.25$C/m$^2$.
Therefore, $P_\mathrm{3d}$ induced by BCL turns out to be typically three orders of magnitude smaller than conventional ferroelectric material.
In addition, if we consider molecular conductors that have larger lattice constant $a \simeq 10$\AA~, the magnitude of the gauge potential $A$ is enhanced ($A=0.5$ for $E=10$MV/cm) and even larger electric polarization could be induced by BCL due to the $A^3$ factor in Eq.~\eqref{Eq:EH_2D_C3}.
For example, the induced electric polarization can be measured as a (pyroelectric) current for a pulsed BCL.
Since the direction of polarization can be controlled with the relative phase $\theta$ of two driving lights, measuring the direction of the BCL induced current that depends on $\theta$ clearly shows the effect of  electric polarization generation by Floquet engineering.

\section*{Acknowledgment}
This work was supported by KAKENHI (20K14407)(SK), JST PRESTO (JPMJPR19L9)(TM), JST CREST (JPMJCR19T3)(SK,TM).


\bibliographystyle{ptephy}
\bibliography{reference}

\appendix

\section{Appendix: Derivation of the effective Hamiltonian for the 2D model \label{app: 2d model}}
Here, we show details of the calculation of the 2D toy model driven by the 3-fold BCL $A_x+iA_y=A_0(e^{i\omega t} + e^{-2i \omega t +i \theta})$ in the Sec.~\ref{subsec: 2d model}. 
The time-dependent Bloch Hamiltonian is given by
\begin{align}
\begin{split}
    H(\vb*{k},t)=
&2t_0 \cos {[ k_x+A_x(t) ]} (\sigma_x \otimes  \sigma_x)
-2\delta t_0 \sin {[ k_x+A_x(t) ]} (\sigma_y \otimes  \sigma_x)\\
+ &2t_0\cos {[ k_y+A_y(t) ]} ( I \otimes  \sigma_x)
+2\delta t_0 \sin {[ k_y+A_y(t) ]} ( \sigma_z \otimes  \sigma_y),
\end{split}
\end{align}
where $\sigma_i$ $(i=x,y,z)$ are Pauli matrices and $\otimes$ denotes tensor product.
Up to the second order of $A_0$, The Fourier coefficients of $H(\vb*{k},t)$ can be calculated as
\begin{align}
  H_0=(2-A_0^2)
\biggl[
    t_0\cos\! k_x (\sigma_x \otimes  \sigma_x)
-\delta t_0 \sin\! k_x (\sigma_y \otimes  \sigma_x)
+ t_0\cos\! k_y ( I \otimes  \sigma_x)
+\delta t_0 \sin\! k_y (\sigma_z \otimes  \sigma_y)
\biggr]  ,
\end{align}
\begin{align}\begin{split}
    H_{\pm 1}=
    t_0 &\biggl[
    -A_0 \sin\! k_x - \frac{1}{2} A_0^2\cos\! k_x  e^{\pm i\theta}    
    \biggr]
    (\sigma_x \otimes  \sigma_x) 
    -\delta t_0 
    \biggl[
    A_0 \cos\! k_x - \frac{1}{2} A_0^2\sin\! k_x  e^{\pm i\theta}    
    \biggr]
    (\sigma_y \otimes  \sigma_x) \\
    +t_0 &\biggl[
    \mp i A_0 \sin\! k_y + \frac{1}{2} A_0^2 \cos\! k_y  e^{\pm i\theta}    
    \biggr]
    ( I \otimes  \sigma_x) 
    + \delta t 
    \biggl[
    \pm i A_0 \cos\! k_y + \frac{1}{2} A_0^2\sin\! k_y  e^{\pm i\theta}    
    \biggr]
    (\sigma_z \otimes  \sigma_y),
\end{split}
\end{align}
\begin{align}
\begin{split}
    H_{\pm 2}=
    t_0 &\biggl[
    - A_0 \sin\! k_x e^{\pm i\theta} - \frac{1}{4} A_0^2 \cos\! k_x      
    \biggr]
    (\sigma_x \otimes  \sigma_x) 
    -\delta t_0 
    \biggl[
    A_0 \cos\! k_x e^{\pm i\theta} - \frac{1}{4} A_0^2 \sin\! k_x     
    \biggr]
    (\sigma_y \otimes  \sigma_x) \\
    +t_0 &\biggl[
    \pm i A_0 \sin\! k_y e^{\pm i\theta} + \frac{1}{4} A_0^2 \cos\! k_y     
    \biggr]
    ( I \otimes  \sigma_x) 
    + \delta t_0 
    \biggl[
    \mp i A_0 \cos\! k_y e^{\pm i\theta} + \frac{1}{4} A_0^2 \sin\! k_y   
    \biggr]
    (\sigma_z \otimes  \sigma_y).
\end{split}
\end{align}
Formulae for commutators of products of Pauli matrices
\begin{align}
   [\sigma_a \otimes \sigma_c, \sigma_b \otimes \sigma_d] &= 2i\epsilon_{abe}\delta_{cd}(\sigma_e\otimes I)+2i\epsilon_{cdf}\delta_{ab}(I \otimes \sigma_f), \\
   [\sigma_a \otimes \sigma_b, I \otimes \sigma_c] &= 2i\epsilon_{bcd}(\sigma_a\otimes \sigma_d),
\end{align}
lead to 
\begin{align}
    [H_{-1},H_1]&=2t_0\delta t_0 A_0^3
\biggl[
    \sin \theta (\sigma_z \otimes  I)
    - \cos \theta (\sigma_z \otimes  \sigma_z)
\biggr]\\
[H_{-2},H_2]&=-t_0\delta t_0 A_0^3
\biggl[
    \sin \theta (\sigma_z \otimes  I)
    - \cos \theta (\sigma_z \otimes  \sigma_z)
\biggr]\\
[H_{-m},H_m]&=0 \ \ (m\geq3).
\end{align}
Up to the first order of $\omega^{-1}$, the effective Hamiltonian 
\begin{align}
    H_{\text{eff}}=H_0+
    \sum_{m\in\mathbb{Z}}\frac{[H_{-m},H_m]}{2m\omega}+\mathcal{O}(\omega^{-2})
\end{align}
is given by
\begin{align}
    H_{\text{eff}}
    =H_0 +
\frac{3}{2\omega}t_0\delta  t_0 A_0^3
\biggl[
    \sin \theta (\sigma_z \otimes  I)
    - \cos \theta (\sigma_z \otimes  \sigma_z)
\biggr] 
+\mathcal{O} (\omega^{-2} , A_0^4).
\end{align}

\section{Appendix: Symmetries of the electric polarization induced by BCL driving \label{app: symmetery polarization}}
In Sec.~\ref{subsec: 2d model}, we briefly explained that the $P_x$-$P_y$ curve reflects the underlying symmetry $C_4$ of the square lattice.
Here, we give the detail of its derivation. Furthermore, we show the relationship between the rose patterns drawn by BCLs and the rotation direction of the electric polarization.

\begin{figure}[t]
    \begin{center}
        \includegraphics[width=\linewidth]{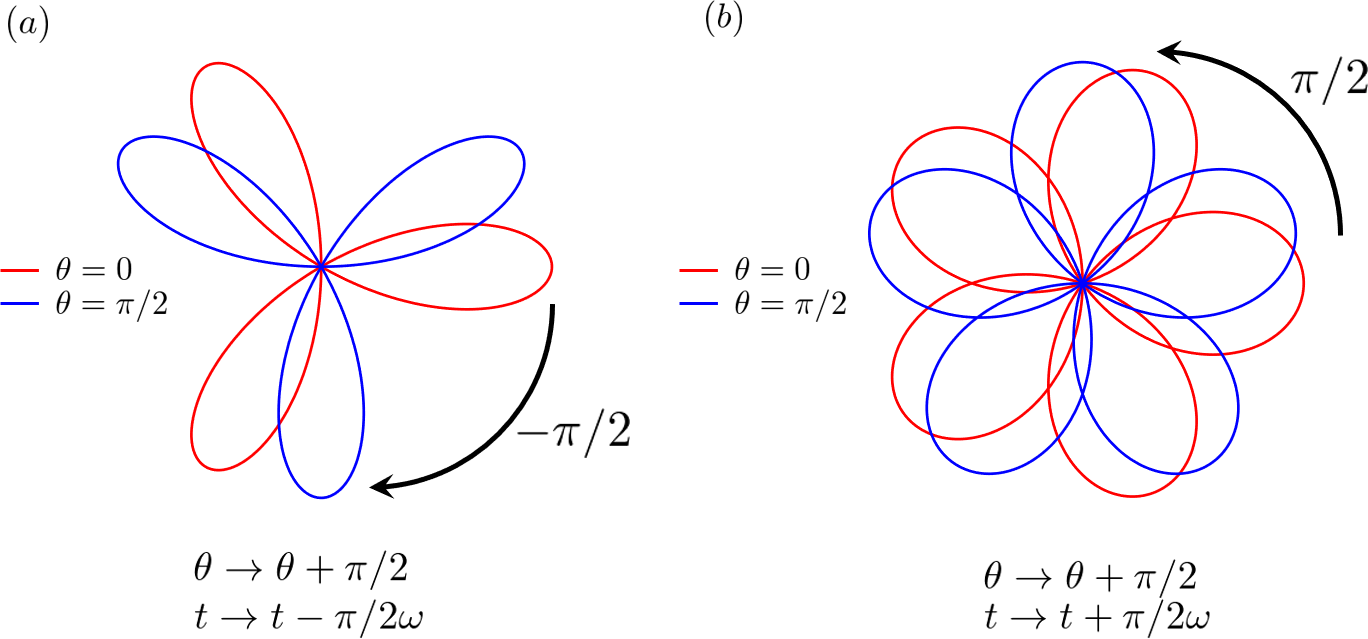}
        \caption{
        Schematic pictures showing that the direction of the rose pattern's rotation by $\theta$-rotation depends on the shape (i.e. fold symmetry) of the rose pattern drawn by BCL. (a) The rose pattern with $C_3$ symmetry rotates counterclockwise by $\pi/2$. (b) The rose pattern with $C_5$ symmetry rotates clockwise by $\pi/2$. 
        }
        \label{figB1}
    \end{center}
\end{figure}

First, in the case of $C_3$ symmetric BCL, the gauge field is given by
$A_x(t,\theta)/A_0=\cos\omega t + \cos (2\omega t -\theta)$ and
$A_y(t,\theta)/A_0=\sin\omega t - \sin (2\omega t -\theta).$
Translation of time $t$ and the relative phase of the two circular lights $\theta$ lead to
\begin{align}\label{Eq:A rotation 1}
    &A_x(t-\pi/2\omega, \theta + \pi/2)=A_y(t,\theta),
    &A_y(t-\pi/2\omega, \theta + \pi/2)=-A_x(t,\theta),
\end{align}
which indicates that $\pi/2$-rotation of $\theta$ and time-translation by $-\pi/2\omega$ lead to $(-\pi/2)$-rotation of $A(t,\theta)$ as illustrated in Fig.~\ref{figB1}(a).
By rotating $\eta=\arg (k_x+ik_y)$ by $-\pi/2$, i.e. $(k_x,k_y)\to (k_y,-k_x)$, in accordance with the above relation,
we obtain
\begin{align}
\begin{split}
    H(k_y,-k_x,t\!-\!\pi/2\omega, \theta \!+\! \pi/2)=
&2t_0 \cos \!{[ k_y\!+\!A_y(t,\theta) ]} (\sigma_x \otimes  \sigma_x)
-2\delta t_0 \sin\! {[ k_y \!+\! A_y(t,\theta) ]} (\sigma_y \otimes  \sigma_x)\\
+ &2t_0\cos\! {[ k_x\!+\!A_x(t,\theta) ]} ( I \otimes  \sigma_x)
-2\delta t_0 \sin\! {[ k_x\!+\!A_x(t,\theta) ]} ( \sigma_z \otimes  \sigma_y).
\end{split}
\end{align}
Thus, we can arrive that this $(-\pi/2)$-rotated Bloch Hamiltonian is unitary equivalent to the original Hamiltonian as
\begin{align}
    U_4^\dagger 
    H(k_y,-k_x,t-\pi/2\omega,\theta+\pi/2)
    U_4
    =H(k_x,k_y,t,\theta),
\end{align}
where the unitary matrix $U_4$ is defined in Eq.~(\ref{Eq:symmetry_system}) and represents the $C_4$ symmetry of the system associated with the rotation of the system by $\pi/2$.
From above, we can conclude that the $C_4$ symmetry of the electric polarization reflects the symmetry respected by the square lattice, i.e.,
\begin{align}\label{Eq:P_symmetry1}
    &P_x(\theta+\pi/2)=P_y(\theta),
    &P_y(\theta+\pi/2)=-P_x(\theta).
\end{align}

In a similar way, we can consider the symmetry property of the $C_5$ symmetric BCL.
When we apply the $C_5$ symmetric BCL
$A_x(t,\theta)/A_0=\cos\omega t + \cos (4\omega t -\theta)$ and
$A_y(t,\theta)/A_0=\sin\omega t - \sin (4\omega t -\theta)$ to the system,
it turns out that
\begin{align}\label{Eq:A rotation 2}
     &A_x(t+\pi/2\omega, \theta + \pi/2)=-A_y(t,\theta),
    &A_y(t+\pi/2\omega, \theta + \pi/2)=A_x(t,\theta),
\end{align}
This shows that $\pi/2$-rotation of $\theta$ and time-translation by $\pi/2\omega$ lead to $\pi/2$-rotation of $A(t,\theta)$, which is the opposite rotation direction from that of the $C_3$ BCL driven case (see Fig.~\ref{figB1}).
In the same way as above, we obtain
\begin{align}
    U_4
    H(-k_y,k_x,t+\pi/2\omega,\theta+\pi/2)
    U_4^\dagger
    =H(k_x,k_y,t,\theta),
\end{align}
which gives another $C_4$ symmetry of the electric polarization
\begin{align}\label{Eq:P_symmetry2}
    &P_x(\theta+\pi/2)=-P_y(\theta),
    &P_y(\theta+\pi/2)=P_x(\theta).
\end{align}
As can be seen from Eq.~(\ref{Eq:P_symmetry1}) and (\ref{Eq:P_symmetry2}), the rotation direction of the electric polarization changes depending on the fold symmetry of the rose pattern drawn by the BCL.

In general, we consider $C_{n+1}$ symmetric BCL given by $A(t,\theta)=A_0(e^{i\omega t}+e^{-in\omega t+i\theta})$, which breaks the $C_4$ symmetry of the square lattice and induces the electric polarization with even integer $n$. Then, we can derive that if $n\equiv 2~(\textrm{mod} 4)$, the electric polarization rotates counterclockwise as in Eq.~(\ref{Eq:P_symmetry1}), and if $n\equiv 0~(\textrm{mod} 4)$, the electric polarization rotates clockwise as in Eq.~(\ref{Eq:P_symmetry2}). This is confirmed from simple relations for an arbitrary integer $m$,
\begin{align}
    &\cos (\omega t -\pi/2) +\cos [(4m+2)(\omega t -\pi/2) - (\theta + \pi /2)]
    =\sin \omega t - \sin  [(4m+2)\omega t  - \theta],\\
    &\sin (\omega t -\pi/2) - \sin [(4m+2)(\omega t -\pi/2) - (\theta + \pi /2)]
    =-\cos \omega t - \cos  [(4m+2)\omega t  - \theta],\\
    &\cos (\omega t +\pi/2) +\cos [4m(\omega t +\pi/2) - (\theta + \pi /2)]
    =-\sin \omega t + \sin  [4m\omega t  - \theta],\\
    &\sin (\omega t +\pi/2) - \sin [4m(\omega t +\pi/2) - (\theta + \pi /2)]
    =\cos \omega t + \cos  [4m\omega t  - \theta].
\end{align}
This obviously shows that Eq.~(\ref{Eq:A rotation 1}-\ref{Eq:P_symmetry1}) are satisfied when $n\equiv 2~(\textrm{mod} 4)$, and Eq.~(\ref{Eq:A rotation 2}-\ref{Eq:P_symmetry2}) are satisfied when $n\equiv 0~(\textrm{mod} 4)$.

\end{document}